\documentclass[12pt]{article}
\newcommand{\bea}{\begin{eqnarray}}
\newcommand{\ena}{\end{eqnarray}}
\usepackage{amsmath,amssymb}
\usepackage{graphicx}
\begin{document}
\topmargin -1cm

\begin{flushright}
\today \\
\end{flushright}

\begin{center}
{\large \bf 
Froggatt-Nielsen hierarchy and the neutrino mass matrix}

\vspace*{10mm}
{\sc Hiroshi Kamikado$^a$}
\footnote{E-mail:kamikado@het.phys.sci.osaka-u.ac.jp},
{\sc Tetsuo Shindou$^{b}$}
\footnote{E-mail: shindou@sissa.it}, 
{\sc Eiichi Takasugi$^{a}$}
\footnote{E-mail: takasugi@het.phys.sci.osaka-u.ac.jp},  
\\
\vspace*{4mm}
${}^{a)}${\em Department of Physics, Osaka University, Toyonaka, Osaka
 560-0043, Japan}\\
${}^{b)}$ {\em Deutsches Elektronen Synchrotron DESY, Notkestrasse 85,
22607 Hamburg, Germany}\\
\end{center}


\begin{abstract}
We study the neutrino mass matrix derived from the 
seesaw mechanism in which the neutrino Yukawa couplings and 
the heavy Majorana neutrino mass matrix are controlled by the Froggatt-Nielsen 
mechanism. In order to obtain the large neutrino mixings, 
two Froggatt-Nielsen fields are introduced with a complex vacuum expectation 
values. As a by-product, CP violation is systematically induced 
even if the order one couplings of FN fields are real. 
We show several predictions of this model, such as
$\theta_{13}$, the Dirac CP phase, two Majorana CP phases, 
the effective mass of the neutrinoless double beta decay and
the leptogenesis. The prediction of the branching 
ratio of $\mu\to e\gamma$ is also given in SUSY model. 

\end{abstract}

\section{Introduction}  

The Froggatt-Nielsen (FN) mechanism\cite{FN} is one of the attractive
mechanisms to explain mass hierarchy of quarks and charged leptons. 
The idea is that the U(1) global symmetry is taken as a flavor symmetry
and the vacuum expectation value (VEV) of a flavor field called FN field
gives a proper structure of Yukawa couplings. 
For quarks and leptons, this mechanism seems to work well by taking 
an appropriate charge assignment of fields. 
However it is known that a mass hierarchy of neutrinos is milder than 
that of charged leptons. 
In the normal hierarchy case,  
$m_2/m_3\simeq \sqrt{\Delta m_{\rm sol}^2/\Delta m_{\rm atm}^2} 
\sim \mathcal{O}(\lambda)$ in contrast to 
$m_{\mu}/m_{\tau}\sim \mathcal{O}(\lambda^2)$. Here $\lambda\sim 0.2$ is 
a size of the Cabbibo angle. 
Then it is a natural question whether the FN mechanism can be applied to 
the mass hierarchy of neutrinos\cite{FNnu}.

A promising approach to get small neutrino masses is 
the seesaw mechanism\cite{seesaw} where the right-handed heavy Majorana 
neutrinos are introduced. The dimension five 
operators are generated after these heavy neutrinos are 
integrated out. This model has several interesting features. 
First of all, the light neutrinos are Majorana particles and we expect 
the neutrinoless double beta decay which 
provides the information not only of light neutrino 
mass scale, but also the Majorana nature of neutrinos such as 
Majorana CP phases\cite{n0bb}. 
Secondly, the scenario of the leptogenesis\cite{leptogenesis} is automatically 
incorporated 
as a mechanism to generate the baryon asymmetry of universe.
In a supersymmetric version of the seesaw model, 
a significant contribution to lepton flavor violation such 
as $\mu\to e\gamma$ can be generated through the running effect 
even if the soft scalar masses are taken as flavor universal 
at the high energy scale\cite{meg}. 
The present bound on 
$\mu\to e\gamma$, $\text{Br}(\mu\to e\gamma)<1.2\times 10^{-11}$
\cite{MEGA} 
will be improved by two order of magnitudes in MEG experiment\cite{MEG-exp} 
and gives the information of neutrino mass matrices. 

Applying the FN mechanism to the neutrino Dirac mass matrix and 
the heavy Majorana neutrino mass matrix would be a natural 
extension of the model. 
Since charged leptons and neutrinos 
are embedded into 
the same SU(2) doublet in left handed sector, a naive expectation 
is that the mass hierarchy in the charged lepton mass matrix and the 
Dirac neutrino mass matrix are similar to each other. In order to 
get the milder hierarchy for neutrino masses, the hierarchy in 
the heavy neutrino mass matrix should compensate 
that of the Dirac mass matrix. 

In this paper, we examine how the milder hierarchy for neutrino masses 
is obtained systematically in the FN scheme. 
We also discuss the problem which is pointed out by 
Koide and Takasugi in the analysis of the 2-3 symmetric 
mass matrices for neutrinos and charged leptons\cite{KoideTakasugi}. 
The 2-3 symmetry\cite{2-3symmetry} is the invariance of the mass 
Lagrangian under the exchange of $\psi_2$ and $\psi_3$, where $\psi_i$ is the 
$i$-th generation fermion of $\psi=e_L, e_R, \nu_L, \nu_R$. 
The problem is as follows: Mass matrices for the neutrino and 
the charged lepton lead to the maximal 2-3 mixing ($\pi/4$ degree 
mixing) for both cases, but 
these mixings cancel each other when they are combined 
to obtain the neutrino mixing. That is, we obtain 
the zero 2-3 mixing. 
This is a serious problem for the 2-3 symmetry, though this is the  
powerful ansatz to restrict the mass Lagrangian. 

We see the general feature of the neutrino mass matrix 
which arises from the seesaw mechanism with the FN mechanism.
We show that the milder hierarchy is naturally obtained 
by a simple ansatz for a choice of the FN 
charge of neutrinos. We discuss a case of real Yukawa coupling constants 
and a case of the mass matrices with the 2-3 symmetry and shows that 
a cancellation occurs for the 2-3 mixing. 
This cancellation can be avoided by considering an
extended FN mechanism \cite{FNCP}, where two FN fields are introduced. 
Quark mass matrices in this type of model has been discussed in 
Ref.\cite{FNCP} 
and it is shown that a CP phase and mixings in 
the Cabbibo-Kobayashi-Maskawa matrix\cite{CKM} are obtained and they 
reproduce the experimental data well.
As a by-product, a CP violation is systematically introduced
by keeping the Yukawa coupling constants for the matter and FN fields 
real. We construct a simple model and examine the light neutrino 
mass hierarchy, 
the CP phases, and the mixings, which  are controlled essentially 
by VEVs of two FN fields. 
We also show several predictions of this model,
$\theta_{13}$, the Dirac CP phase, two Majorana CP phases, 
the effective mass of the neutrinoless double beta decay and
the leptogenesis. The prediction of the branching 
ratio of $\mu\to e\gamma$ is also given in SUSY model.
 
\section{The neutrino mass matrix in the hierarchy scheme}

We assume that the hierarchy of mass matrices arise from 
the Froggatt-Nielsen (FN) nonrenormalizable interaction as
\begin{eqnarray}
\cal{L}_{FN} &=&
 - \overline{\nu_{Ri}} (Y)_{ij} \ell_j \cdot H_u 
 \left( \frac{\Theta}{\Lambda} 
\right)^{f_{\nu_{Ri}} + f_{\ell_j}}\nonumber\\
&& - \overline{e_{Ri}} (Y_e)_{ij} \ell_j \cdot H_d 
\left( \frac{\Theta}{\Lambda} \right)^{f_{E_{Ri}} + f_{\ell_j}},
\nonumber\\
&& -\frac12 \bar{\nu_{Ri}}m_{ij}(\nu_{Rj})^c
\left( \frac{\Theta}{\Lambda} \right)^{f_{\nu_{Ri}} + f_{\nu_{Rj}} },
\end{eqnarray}
where $H_{u,d}$ are Higgs doublets, $L_j$ is the 
left-handed lepton doublets, $E_{Rj}$ and $\nu_{Rj}$ are right-handed 
charged leptons and right-handed neutrinos in the j-th generation, 
respectively. 
$\Lambda$ is a cut-off scale, $Y_\nu$ and $Y_e$ are coupling constants, 
and their U(1) charges are expressed by $f$. By assigning the U(1) 
charge of $\Theta$ by $f_\Theta=-1$ and $f_{H_u}=f_{H_d}=0$. 
The Lagrangian is invariant under $U(1)$ transformation. 

When $\Theta$ takes a vacuum expectation value 
\begin{eqnarray}
\langle\Theta\rangle=\lambda \Lambda \;, 
\end{eqnarray}
where $\lambda$ is a small quantity which is of order of Cabbibo angle, 
$\sim 0.2$. 
Then we obtain effective Yukawa couplings matrices and the right-handed 
neutrino mass matrix as
\begin{eqnarray}
(Y^{\rm eff})_{ij} &=& 
(Y)_{ij} \lambda^{f_{\nu_{Ri}} + f_{L_j} }.\nonumber\\
(Y_e^{\rm eff})_{ij} &=& 
(Y_e)_{ij} \lambda^{f_{e_{Ri}} + f_{L_j} + f_{H_{d}}}.\nonumber\\
(m^{\rm eff})_{ij} &=& k_M
m_{ij} \lambda^{f_{\nu_{Ri}} + f_{\nu_{Rj}} }.
\end{eqnarray}
Since the power of $\lambda$ provides the hierarchical structure of 
mass matrices, coupling constants, $(Y_\nu)_{ij}$ are approximately 
equal each other. This is true for  $(Y_e)_{ij}$ and $m_{ij}$.   

Let us consider neutrino mass matrices. For simplicity, we express 
FN charges for left-handed lepton doublets and right-handed neutrinos as 
$(f_{L_1},f_{L_2},f_{L_3})
=(f_1,f_2,f_3)$ and $(f_{\nu_{R1}},f_{\nu_{R 2}},f_{\nu_{R 3}})
=(g_1,g_2,g_3)$. Then, the Dirac and Majorana neutrino mass matrices are 
given by
\bea
m_D=v_u k_\nu \begin{pmatrix}
a_{11}\lambda^{g_1+f_1}& a_{12}\lambda^{g_1+f_2}& a_{13}\lambda^{g_1+f_3}\cr
a_{21}\lambda^{g_2+f_1}& a_{22}\lambda^{g_2+f_2}& a_{23}\lambda^{g_2+f_3}\cr
a_{31}\lambda^{g_3+f_1}& a_{32}\lambda^{g_3+f_2}& a_{33}\lambda^{g_3+f_3}\cr
\end{pmatrix}
\ena
\bea
M_R=m\begin{pmatrix}
b_{11}\lambda^{2g_1}&b_{12}\lambda^{g_1+g_2}&b_{13}\lambda^{g_1+g_3}\cr
b_{12}\lambda^{g_2+g_1}&b_{22}\lambda^{2g_2}&b_{23}\lambda^{g_2+g_3}\cr
b_{13}\lambda^{g_3+g_1}&b_{23}\lambda^{g_3+g_2}&b_{33}\lambda^{2g_3}\cr
\end{pmatrix}
\ena
where $k_\nu a_{ij}=(Y)_{ij}$ and $m b_{ij}=m_{ij}$, so that 
$a_{ij}$ and $b_{ij}$ are normalized to  quantities of order 1. 
 
Now, by the seesaw mechanism, the neutrino mass is given by 
\bea
m_\nu=m_D^TM_R^{-1}m_D =\frac{(v_uk_\nu)^2}{m}
\begin{pmatrix}
D\lambda^{2(f_1-f_3)}&A\lambda^{f_1+f_2-2f_3}&A'\lambda^{f_1-f_3}\cr
A\lambda^{f_1+f_2-2f_3)}&B\lambda^{2(f_2-f_3)}&C\lambda^{f_2-f_3}\cr
A'\lambda^{f_1-f_3}&C\lambda^{f_2-f_3}&B'\cr
\end{pmatrix}\;,
\ena
where $A$, $A'$, $B$, $B'$, $C$ and $D$ are functions of $a_{ij}$ and 
$B_{ij}$ and do not contain $\lambda$. 
From Eq.(6), we observe an interesting features: 
(a) The hierarchical structure of neutrino mass matrix depends only on 
the FN charge of the left-handed neutrinos, $f_i$ and independent on 
those of the right-handed neutrinos, $g_i$  
(b) If we take $f_2=f_3=f_1+1$ which is reasonable in view of SU(5) 
GUT where the FN charges of $d_{Ri}$ quarks are taken as 
$(2,1,1)$ and $\nu_{Li}$ and $d_{Ri}$ form a same multiplet. 

By taking $f_2=f_3=f_1+1$, we obtain 
\bea
m_\nu =\frac{(v_uk_\nu)^2\lambda^{2f_3}}{m}
\begin{pmatrix}
D\lambda^{2}&A\lambda &A'\lambda\cr
A\lambda&B&C\cr
A'\lambda&C&B'\cr
\end{pmatrix}\;.
\ena
Needless to say that parameters $A$, $B$, $B'$, $C$, $D$ are expected to 
be quantities of order one. 

This matrix leads an interesting neutrino mass patterns and mixings, 
the large 2-3 mixing, the reasonably large 1-2 mixing, and the 
mass squared ratio. In order to see the neutrino mixings, we have to 
take into account the charged lepton mass matrix, 
because the neutrino mixing matrix is obtained by multiplying the 
matrices which diagonalize the charged lepton mass matrix and the 
neutrino mass matrix. 

How about the charged lepton mass matrix? 
Since $e_{Li}$ and $\nu_{Li}$ form a doublet of the electroweak symmetry, 
the FN charges of $e_{Li}$ should be the same as those of $\nu_{Li}$. 
By assuming the FN charges of $e_{Ri}$ as $(k_1,k_2,k_3)$ 
$(k_1>k_2>k_3)$, we find
\bea
m_e=v_d k_e \begin{pmatrix}
c_{11}\lambda^{k_1+f+1}& c_{12}\lambda^{k_1+f}& c_{13}\lambda^{k_1+f}\cr
c_{21}\lambda^{k_2+f+1}& c_{22}\lambda^{k_2+f}& c_{23}\lambda^{k_2+f}\cr
c_{31}\lambda^{k_3+f+1}& c_{32}\lambda^{k_3+f}& c_{33}\lambda^{k_3+f}\cr
\end{pmatrix}\;.
\ena
Then, we find
\bea
m_e^\dagger m_e =(v_dk_e)^2 \lambda^{2k_3+2f}
 \begin{pmatrix}
D_e\lambda^{2}& A_e\lambda& {A'}_e\lambda\cr
A_e^*\lambda&B_e& C_e\cr
{A'}_e^*\lambda&C_e^*& B_e'\cr
\end{pmatrix}\;.
\ena
Similarly to the neutrino mass matrix case, the same hierarchy is 
obtained as the neutrino mass matrix due to the fact that the 
FN charges for $e_{Li}$ are the same as those for $\nu_{Li}$.

\noindent
(a) The case of real coupling parameters 

We  consider the case 
where all coupling constants, $Y_{ij}$, $(Y_{e})_{ij}$ and $m_{ij}$ are 
real. As we saw, both the neutrino mass matrix and the charged lepton 
mass matrix lead large 2-3 mixing. 
The large 2-3 mixing given by the neutrino tends to be cancelled 
by the large 2-3 mixing given by the charged lepton, so that 
the resultant atmospheric neutrino mixing, i.e., 
the mixing between $\nu_\mu$ and $\nu_\tau$ mixing is small. 
This may be a somewhat generic and serious problem. 

\noindent
(b) The case of complex coupling parameters

If we consider complex coupling parameters, there appear too many 
free parameters including phases, so that the model 
loses the predictive power. In order to decrease the freedom and make 
the model predictive, the 2-3 symmetry\cite{KoideTakasugi,2-3symmetry} is frequently used, 
where the 2-3 symmetry requires the invariance under the exchange of 
the 2nd and the 3rd generations. If we apply this symmetry for the neutrinos, 
by assuming $g_3=g_3$ in addition to $f_2=f_3$,  we find 
$A=A'$ and $B=B'$. Then, 
\bea
\begin{pmatrix}
1&0&0 \cr
0& \frac{1}{\sqrt 2}&\frac{1}{\sqrt 2}\cr
0&-\frac{1}{\sqrt 2}&\frac{1}{\sqrt 2}\cr
\end{pmatrix}
\begin{pmatrix}
D\lambda^{2}&A\lambda&A\lambda\cr
A\lambda&B&C\cr
A\lambda&C&B\cr
\end{pmatrix}
\begin{pmatrix}
1&0&0 \cr
0& \frac{1}{\sqrt 2}&-\frac{1}{\sqrt 2}\cr
0&\frac{1}{\sqrt 2}&\frac{1}{\sqrt 2}\cr
\end{pmatrix}
=
\begin{pmatrix}
D\lambda^{2}&\sqrt{2} A\lambda&0\cr
\sqrt{2}A\lambda&B+C&0\cr
0&0&B-C\cr
\end{pmatrix}\;.
\ena
If we take $|B-C|\gg|B+C|\sim A\lambda$, the mass spectrum is 
\bea
m_3&=&B-C\;,\nonumber\\
m_2&\simeq& B+C+\frac{2A^2}{B+C}\lambda^2\;,\nonumber\\
m_1&\simeq& -\frac{2A^2}{B+C}\lambda^2
\ena
and thus
\bea
\frac{m_2^2-m_1^2}{m_3^2}\simeq \frac{(B+C)^2+4A^2\lambda^2}{(B-C)^2}
\ena
which is a quantity of order $\lambda^2\sim 0.04$ and close to 
$\Delta m^2_{sol}/\Delta m^2_{atom}\sim 0.03$. 
If we take $(B+C)\sim A\lambda$, which is consistent with the assumed 
size for the mass squared ratio, then, mass matrix for the 1st and 2nd 
columns becomes 
\bea
\sim A\lambda \begin{pmatrix}
0&\sqrt{2}\cr
\sqrt{ 2}&1\cr
\end{pmatrix}
\ena
and lead to the solar mixing of 
$\tan^2 \theta_{12}\simeq 1/2$. 

For the charged lepton mass matrix, we have two cases. One is the case 
where $k_2\neq k_3$. Then, we only require the 2-3 symmetry for $e_L$. 
In this case, we find the relations $c_{i2}=c_{i3}$, which lead to 
$A_e=A_e'$, $B_e=B_e'=C_e=C_e^*$. Another one is to assume the 2-3 symmetry 
for $e_R$ also by taking $k_2=k_3$. In this case, we find 
$A_e=A_e'$, $B_e=B_e'$ and real $C_e$. For both cases, the matrix is 
block diagonalized by the rotation with the angle 
of $\pi/4$ as in the case of Eq.(10), which cancels the $\pi/4$ mixing 
which came from the diagonalization of the neutrino mass matrix. 
As a result, the atmospheric neutrino mixing vanishes. 

As we saw, the FN type hierarchical model leads either 
the small atmospheric neutrino mixing, or the loss of the predictive 
power. 
In the next section, we give a way to avoid this problem 
in the two FN fields model, where the relative 
phase of their vacuum expectation values works to lead the 
mismatch and also to introduce the 
CP violation, while coupling parameters are taken to be real.




\section{ The model of neutrino mixing in the extended Froggatt-Nielsen 
mechanism}

We consider a model which consists of two FN fields, $\Theta_{1}$, $\Theta_{2}$ and assume 
that $Y_\nu$, $Y_e$ and $m$ are real matrices. In this scheme 
the CP violation is originated solely from the relative phase 
of vacuum expectation values of two FN fields. As it was shown in Ref.~\cite{FNCP},
we have to introduce $Z_2$ symmetry in order for 
this phase to work as the Dirac CP phase. We take the FN charge and 
$Z_2$ parity for them as
\begin{eqnarray}
(f_{\Theta_1},f_{\Theta_2},P_{\Theta_1},P_{\Theta_2})=(-1,-1,+,-)\;,
\end{eqnarray}
where $P_{\Theta_i}$ gives the $Z_2$ parity. 
For leptons, we take 
\bea
&&(f_{\ell_1}, f_{\ell_2}, f_{\ell_3},P_{\ell_1}, P_{\ell_2}, P_{\ell_3})
=(2,1,1,+,+,-),\nonumber\\
&&(f_{e_R 1}, f_{e_R 2}, f_{e_R 3},P_{e_R1}, P_{e_R2}, P_{e_R3})=
(3,2,0,+,+,-),\nonumber\\
&&(f_{\nu_R 1}, f_{\nu_R 2}, f_{\nu_R 3},P_{\nu_R1}, P_{\nu_R2}, P_{\nu_R3})
=(2,1,0,+,+,-).
\ena
and
for Higgs, we take $(f_{Hu},f_{Hd},P_{Hu},P_{Hd})=(0,0,+,+)$. This choice of 
the FN charges are consistent to that of quarks in the SU(5) GUT scheme. 
The interaction Lagrangian is given by,
\begin{eqnarray}
\cal{L_{FN \rm{2}}} &=& - \sum_
{n_{\nu 1},n_{\nu 2}} \overline{\nu_{Ri} }
(Y)_{ij} \ell_j \cdot H_u 
\left( \frac{\Theta_1}{\Lambda} \right)^{n_{\nu 1}} 
\left( \frac{\Theta_2}{\Lambda} 
\right)^{n_{\nu 2}} 
\nonumber\\
&&- \sum_{n_{e 1},n_{e2}} \overline{e_{Ri}}(Y_e)_{ij} \ell_j \cdot H_d 
\left( \frac{\Theta_1}{\Lambda} \right)^{n_{e1}} 
\left( \frac{\Theta_2}{\Lambda} 
\right)^{n_{e2}} \;,\nonumber\\
&&- \sum_{n_{M1},n_{M2}} \overline{\nu_{Ri}}(m)_{ij} (\nu_{Rj})^C 
\left( \frac{\Theta_1}{\Lambda} \right)^{n_{M1}} 
\left( \frac{\Theta_2}{\Lambda} 
\right)^{n_{M2}} \\
\;.
\end{eqnarray}
where $(n_{X1},n_{X2})$ are taken so as to keep the invariance of 
the FN U(1) symmetry and $Z_2$ symmetry. 
Effective Yukawa couplings and the heavy neutrino masses are 
given by 
\begin{eqnarray}
{\cal{L_{\rm eff}}} =-\overline{\nu_{Ri} }(y)_{ij} \ell_j \cdot H_u 
-\overline{e_{Ri}}(y_e)_{ij} \ell_j \cdot H_d 
- \overline{\nu_{Ri}}(M_R)_{ij} (\nu_{Rj})^C \;
\end{eqnarray}
are given by taking 
\begin{eqnarray}
\lambda=\frac{\langle\Theta_1\rangle}{\Lambda}\;,\;
R=\frac{\langle\Theta_2\rangle}{\langle\Theta_1\rangle}\equiv 
|R| e^{i \alpha}\;,
\end{eqnarray}

\vskip 3mm
\noindent
(a) A model of mass matrices

By taking $\langle H_u\rangle=v_u$ and  $\langle H_d\rangle=v_d$, mass matrices are obtained. 
As we stated before, in the spirit of 
the FN hierarchy model, elements of $Y$ are considered to be approximately 
equal each other, and for $Y_e$ and $m$, this should hold. 
In the following, we  
consider a simple model where $Y$, $Y_e$ and $m$ are proportional to 
the democratic matrix as
\bea
Y=k_\nu \begin{pmatrix}
1&1&1 \cr
1&1&1\cr
1&1&1\cr
\end{pmatrix}\;,
Y_e=k_e \begin{pmatrix}
1&1&1 \cr
1&1&1\cr
1&1&1\cr
\end{pmatrix}\;,
m=m_M \begin{pmatrix}
1&1&1 \cr
1&1&1\cr
1&1&1\cr
\end{pmatrix}\;.
\ena
Then, we find
\bea
m_D&=&v_u y=v_u k_\nu
\begin{pmatrix}
B_4\lambda^4&B_2\lambda^3&RB_2\lambda^3\\
B_2\lambda &B_2\lambda^2&R\lambda^2\\
R\lambda^2&R\lambda&\lambda
\end{pmatrix}\;,\nonumber\\
M_e&=&v_d k_e
\begin{pmatrix}
B_4\lambda^5&B_4\lambda^4&RB_2\lambda^4\\
B_4\lambda^4&B_2\lambda^3&RB_2\lambda^3\\
R\lambda^2&R\lambda&\lambda
\end{pmatrix}\;, \nonumber\\
M_R&=&m_M
\begin{pmatrix}
B_4\lambda^4&B_2\lambda^3&R\lambda^2\\
B_2\lambda^3&B_2\lambda^2&R\lambda\\
R\lambda^2&R\lambda&1
\end{pmatrix}
\;,
\ena
where $B_{2n}=1+R^2+\cdots +R^{2n}$. 
We are interested in whether this simple model can reproduce 
the observed data. Firstly, we derive the neutrino mass matrix for 
the left-handed neutrinos,
\bea
m_\nu=m_D^TM^{-1}m_D= v_L\begin{pmatrix}
B_4\lambda^2&B_2\lambda&RB_2\lambda\\
B_2\lambda^3&B_2&R\\
RB_2\lambda&R&B_2
\end{pmatrix},
\ena
where $v_L=(v_u k_\nu)^2\lambda^2/m_M$.

\vskip 3mm
\noindent
(b) Diagonalization

At first, we discuss the neutrino mass diagonalization. 
By the transformation of the unitary matrix $U^{(1)}$, where
\bea
U^{(1)}=\begin{pmatrix}
1&0&0 \cr
0& \frac{1}{\sqrt 2}&\frac{1}{\sqrt 2}\cr
0&-\frac{1}{\sqrt 2}&\frac{1}{\sqrt 2}\cr
\end{pmatrix}
\begin{pmatrix}
1&0&\frac{(1+R)^*B_2^*}{\sqrt2 (B_2+R)^*}\lambda \cr
0& 1&0\cr
-\frac{(1+R)B_2}{\sqrt2 (B_2+R)}\lambda&0&1\cr
\end{pmatrix}\;,
\ena  
$m_\nu$ is block diagonalized in a good approximation as
\bea
(U^{(1)})^Tm_\nu U^{(1)}=v_L\begin{pmatrix}
(B_4-\frac{((1+R)B_2)^2}{2(B_2+R)})\lambda^2
&\frac{(1-R)B_2}{\sqrt 2}\lambda&0 \cr
\frac{(1-R)B_2}{\sqrt 2}\lambda
& B_2-R&0\cr
0&0&B_2+R\cr
\end{pmatrix}\;,
\ena  
where we assumed that $|B_2+R|\gg|B_2-R|$. 
The matrix in Eq.(21) is diagonalized by the matrix $U^{(3)}$
\bea
U^{(2)}=\begin{pmatrix}
c&-se^{i\rho}&0\cr
se^{i\rho}&ce^{2i\rho}&0\cr
0&0&1\cr
\end{pmatrix}\;,
\ena
where $c=\cos \theta$ and $s=\sin \theta$, and 
\bea
\rho&=&{\rm arg}\left( \frac{(1-R)B_2}{B_2-R}\right)+\pi\;,\nonumber\\
\tan 2\theta &=&{\sqrt2} \left|\frac{(1-R)B_2}{B_2-R}\right| \lambda\;,
\ena
and neutrino masses are
\bea
m_1&\simeq&-v_L[s^2-{\sqrt2}sc |\frac{(1-R)B_2}{B_2-R}|]
|B_2-R| e^{i\alpha_1}=v_L\frac{s^2}{c^2-s^2}|B_2-R| e^{i\alpha_1}\;,
\nonumber\\
m_2&\simeq&v_L[c^2+{\sqrt2}sc |\frac{(1-R)B_2}{B_2-R}|]|B_2-R|
 e^{i\alpha_2}=v_L\frac{c^2}{c^2-s^2}|B_2-R| e^{i\alpha_2}\;,\nonumber\\
m_3&\simeq &v_L|B_2+R|e^{i\alpha_3}\;, 
\ena
where $\cos 2\theta=c^2-s^2>0$ is taken and 
\bea
\alpha_1&=& 2\rho+{\rm arg}(B_2-R)+\pi\;,\nonumber\\
\alpha_2&=& 4\rho+{\rm arg}(B_2-R)\;,\nonumber\\
\alpha_3&=& {\rm arg}(B_2+R)\;.
\ena
Now, we define the phase matrix 
\bea
P={\rm diag}(e^{-i\frac{\alpha_1}2},
e^{-i\frac{\alpha_2}2},e^{-i\frac{\alpha_3}2})\;,
\ena
the matrix which diagonalizes the 
neutrino mass matrix is $U_\nu=U^{(1)}U^{(2)}P$.

Next, we go to the diagonalization of the charged lepton mass matrix. 
By transforming $M_e$ as $V_e^\dagger M_e U_e$ by 
the unitary matrix, $U_e=V^{(1)}V^{(2)}$, where
\bea
V^{(1)}&=&\begin{pmatrix}
1&0&0 \cr
0& \frac{1}{\sqrt{ 1+|R|^2}}&\frac{R^*}{\sqrt{1+|R|^2}}\cr
0&-\frac{R}{\sqrt{1+|R|^2}}&\frac{1}{\sqrt{1+|R|^2}}\cr
\end{pmatrix}
\begin{pmatrix}
1&0&\frac{R^*}{\sqrt{1+|R|^2}}\lambda \cr
0& 1&0\cr
-\frac{R}{\sqrt{1+|R|^2}}\lambda&0&1\cr
\end{pmatrix}\;,\nonumber\\
V^{(2)}&=&\begin{pmatrix}
1&\frac{1+|R|^2R^{*4}}{(\sqrt{1+|R|^2})^{3/2}}\lambda&0 \cr
-\frac{1+|R|^2R^{4}}{(\sqrt{1+|R|^2})^{3/2}}\lambda&1&0\cr
0& 0&1\cr
\end{pmatrix}\;
\ena
$M_e$ is diagonalized in a good approximation and the eigenvalues 
are
\bea
\frac{m_\mu}{m_\tau}=\frac{|1-R^4|}{1+|R|^2}\lambda\;,
\frac{m_e}{m_\mu}=\frac{|R|^4{\sqrt{1+|R|^2}}}{|1-R^4|^2}\lambda^2\;.
\ena

\vskip 3mm
\noindent
\section{The neutrino mixing matrix and mass spectrum of leptons}
At first, we notice that the neutrino mixing which is given by
\bea
V=U_e^\dagger U_\nu=V^{(2)\dagger} V^{(1)\dagger}U^{(1)}U^{(2)}P\;,
\ena
and consider how the cancellation of the 2-3 mixing is avoided. 
The 2-3 mixing is essentially given in the following part
\bea
\begin{pmatrix}
1&0&0 \cr
0& \frac{1}{\sqrt{1+|R|^2}}&-\frac{R^*}{\sqrt{1+|R|^2}}\cr
0&\frac{R}{\sqrt{1+|R|^2}}&\frac{1}{\sqrt{1+|R|^2}}\cr
\end{pmatrix}
\begin{pmatrix}
1&0&0 \cr
0& \frac{1}{\sqrt 2}&\frac{1}{\sqrt 2}\cr
0&-\frac{1}{\sqrt 2}&\frac{1}{\sqrt 2}\cr
\end{pmatrix}
=\begin{pmatrix}
1&0&0 \cr
0& \frac{1+R^*}{\sqrt{2(1+|R|^2)}}&\frac{1-R^*}{\sqrt{2(1+|R|^2)}}\cr
0&-\frac{1-R}{\sqrt{2(1+|R|^2)}}&\frac{1+R}{\sqrt{2(1+|R|^2)}}\cr
\end{pmatrix}\;.
\ena
Now, the atmospheric mixing is 
\bea
\sin^2 2\theta_{\rm atm}\simeq \frac{|1-R^2|^2}{(1+|R|^2)^2}
= 1-\left(\frac{2\cos \alpha}{|R|+\frac1{|R|}}\right)^2\;.
\ena
If there is no phase in $R$, i.e., $\alpha=0$, then the  2-3 mixings 
for the charged lepton and the neutrino cancel each other and 
leads to the small mixing because we consider $|R|\simeq 1$. 
Even in the existence of the phase $\alpha$, it is hard to achieve 
$\sin^2 \theta_{\rm atm}=1$. Here, we relax this condition and 
want to reproduce $\sin^2 \theta_{\rm atm}\ge 0.9$, then we find 
\bea
|\cos \alpha|\le \frac{|R|+\frac1{|R|}}{2\sqrt{10}}\;.
\ena
Since we expect $|R|\simeq 1$, $|\cos \alpha|$ must be 
around $1/\sqrt{10}$. 

Next, we impose the condition which assures the computation given 
above. The condition is 
$|B_2+R|\gg|B_2-R|$ which is needed to get the hierarchy of 
neutrino mass $|m_3|\gg|m_2|$. This requires that
\bea
\frac{((|R|+\frac1{|R|})\cos \alpha -1)^2+(|R|-\frac1{|R|})^2\sin^2 \alpha}
{((|R|+\frac1{|R|})\cos \alpha +1)^2+(|R|-\frac1{|R|})^2\sin^2 \alpha}
\ll 1\;.
\ena
To fulfill this condition with $\cos \alpha =1/\sqrt 10$, 
we need $(|R|+\frac1{|R|})\cos \alpha \sim 1$, so that 
$|R|\sim 1/|R|$ is required There is another reason. Let us see the 
11 element of the neutrino mixing by neglecting $O(\lambda^2)$ term,
\bea
V_{11}=\left[c-se^{i\rho}\frac1{{\sqrt 2}(1+|R|^2)}\left(
\frac{1+|R|^2R^{*4}}{(1-R^*)B_2^*}-(1-R)R^*\right)\lambda\right]
e^{-i\frac{\alpha_1}2}\;.
\ena
Here, we observe that the 2nd term in the parenthesis becomes real in the 
limit of $|R|=1$ and works to cancel the first term $c$ 
when $\cos \alpha=\frac1{\sqrt 10}$. This means that we obtain  
smaller $|V_{11}|$ element, which in turn leads to 
a larger solar mixing angle. 

Before going into the detailed analysis, we comment about the 
number of parameters. There are three parameters, $\lambda$, $|R|$ and 
$\alpha$ aside from the overall normalization of $m_D$, $M_R$ and 
$m_\ell$. Therefore, if we fix the above three parameters, three 
neutrino mixings, one Dirac CP phase, two Majorana CP phases, 
the ratios of masses of the left-handed neutrinos and the right-handed 
neutrinos, and those 
of charged leptons. If we go to the leptogenesis and the 
LFV, we need to fix the overall factors, which we see later. 
 
In the following analysis, we take the values of three parameters as
\bea
\lambda=\frac14\;,\;\;\cos \alpha=\frac1{\sqrt{10}}\;, 
\sin \alpha=-\frac3{\sqrt{10}}\;, 
\;\;|R|=1\;,
\ena
as we explained before. 
Then, we find
\bea
V=P'\begin{pmatrix}
0.863&0.585&(-0.012+0193i)\cr
(-0.531-0.054i)&(0.678+0.069i)&0.585\cr
(0.259+0.026i)&(-0.528-0.054i)&0.811\cr
\end{pmatrix}
\begin{pmatrix}
1&0&0\cr
0&e^{-0.50\pi i}&0\cr
0&0&e^{0.53\pi i}\cr
\end{pmatrix}\;,
\ena
where $P'$ is a physically meaningless diagonal phase matrix. 
We see the Dirac phase $\delta=-0.52\pi$ in the
standard phase convention\cite{PDGphase}
and two Majorana phases as $\beta=-0.50\pi$ and $\gamma=0.53\pi$.

We obtain the ratios of masses as
\bea
\frac{m_1}{m_3}=0.027\;,&&
\frac{m_2}{m_3}= 0.27 \nonumber\\
\frac{M_1}{M_3}=0.0035\:,&&
\frac{M_2}{M_3}=0.060\nonumber\\
\frac{m_e}{m_\tau}=0.00062\;,&&
\frac{m_\mu}{m_\tau}=0.037\;.
\ena
The absolute values of elements of the mixing matrix  are
\bea
\begin{pmatrix}
0.863&0.585&0.194\cr
0.533&0.681&0.585\cr
0.260&0.531&0.811\cr
\end{pmatrix}
\ena
which gives
\bea
\tan^2 \theta_{\rm sol}=0.455\;, \;\;\sin^2 2\theta_{\rm atm}=0.9\;,
\ena
which are in a good agreement with the data, and 
\bea
\frac{\Delta m_{\rm sol}}{\Delta m_{\rm atm}}&\simeq &
\frac{m_2^2-m_1^2}{m_3^2}=6.3\times 10^{-2}\;, 
\ena
which are in a reasonable agreement with the experimental data, 
in view of this simple model. 

By taking $m_\tau=1.75$GeV and $m_{\nu \tau}=m_{\nu 3}\simeq
\sqrt{\Delta m_{\rm atm}^2}\simeq 0.05$eV, we find 
$m_e= 1.1$MeV and $m_\mu=64$MeV, and 
$m_{\nu e}=m_{\nu 1}=\mathcal{O}(10^-3)$eV and 
$m_{\nu \mu}=m_{\nu 2}=0.012$eV. 
In order to obtain $m_{\nu_3}=0.05$eV, 
$v_L= 3.0\times 10^{-11}$GeV is required.
Though the predicted masses in our very simple model don't have
excellent agreements with the experimental data, {\it i.e.}
the predicted value of $m_e$ is about twice of the experimental 
value and that of $m_\mu$ is  about half, we may say that our model 
is still successful.
Such a discrepancy can be improved by relaxing our assumption that 
$Y$, $Y_e$ and $m$ are proportional to the democratic matrix. 

We comment that if $\sin \alpha=\frac3{\sqrt{10}}$ is taken, then
the mixing matrix $V$ changes to $V^*$ in comparison with the case of 
$\sin \alpha=-\frac3{\sqrt{10}}$.

\section{Predictions}

Since we fixed all parameters except for $m_M$ which determines 
the absolute magnitude of the right-handed neutrino masses, we 
can compute predictions for various observables. 

\vskip 3mm
\noindent
(a) The neutrinoless double beta decay 

As a feasible experiment which has a potential to 
give an information of Majorana CP violation phases 
at low energy scale,  we consider 
the neutrinoless double beta decay \cite{n0bb}, 
$(A,Z) \rightarrow (A,Z+2) + e^- + e^-$. 
If the neutrinoless double beta decay is observed, 
the measurement of the neutrinoless double beta decay half-life
combined with information on the absolute values of neutrino masses 
might give a constraint on the neutrino mass parameters or determine 
them. 
Predictions on the neutrino less double beta decay can be controlled 
by a effective mass, $\langle m_\nu\rangle=|\sum V_{ej}^2 m_j|$\cite{n0bb}.
In our model, the effective mass is predicted as
\begin{align}
\frac{\langle m_\nu\rangle}{m_3}=0.03\;.
\end{align}
Therefore, with $m_3=\sqrt{\Delta m^2_{atm}}\sim 0.05$eV, we find 
that the effective mass is as small as 0.0015eV, 
so that it seems hard to be observed.

\vskip3mm
\noindent
(b) Baryon number asymmetry

We consider the thermal leptogenesis scenario\cite{leptogenesis} 
in which the baryon number asymmetry of the universe is generated by 
the conversion of  
the Lepton number asymmetry produced by CP violating decay of
heavy right-handed neutrinos. 
Recently the effect of flavors is  
studied \cite{FlvLepto,FlvLepto-2}, and this effect is 
shown to be significant 
in several cases,
{\it e.g.}, a case that 
primordial $B-L$ asymmetries are considered\cite{Primodial} or 
one that total CP asymmetry parameter is strongly suppressed by the 
cancellation between the flavor dependent CP asymmetries as
$\epsilon_1^e+\epsilon_1^{\mu}+\epsilon_1^{\tau}\sim 0$\cite{zero-eps}. 
However, in most cases, the contribution from flavor effect is within 
10\%\cite{FlvLepto-2}. 
Here we consider the zero primordial asymmetry case and the total CP 
asymmetry parameter is not small. Then one flavor 
approximation can be used. 

Since the right-handed neutrino mass eigenvalues are hierarchical, 
the CP asymmetry parameter and washout mass parameter in our model 
with $\lambda = 1/4$, $|R|=1$ and $\cos\alpha =1/\sqrt{10}$, 
$\sin\alpha = -3/\sqrt{10}$ 
for the standard model (SM) case and 
minimal supersymmetric standard model (MSSM) case
are
\begin{align}
\epsilon_1 \equiv &
\frac{\Gamma(\nu_{R1} \to l H_u)-\Gamma(\nu_{R1}\to l^c H_u^*)}
{\Gamma(\nu_{R1} \to l H_u)+\Gamma(\nu_{R1}\to l^c H_u^*)}\nonumber\\
\sim&
\begin{cases}
-\frac{3}{16\pi (YY^{\dagger})_{11}}
\sum_{j\neq 1}\frac{M_1}{M_j}\mathrm{Im}\left((YY^{\dag})_{j1}^2\right)
\simeq -1.3\times 10^{-6}\left(\frac{m_M}{10^{13}\text{GeV}}\right)\;,
&\text{SM case}\;,\\
-\frac{1}{8\pi (YY^{\dagger})_{11}}
\sum_{j\neq 1}\frac{M_1}{M_j}\mathrm{Im}\left((YY^{\dag})_{j1}^2\right)
\simeq -8.3\times 10^{-7}\left(\frac{m_M}{10^{13}\text{GeV}}\right)\;,
&\text{MSSM case}\;,
\end{cases}
\end{align}
and
\begin{align}
\tilde{m}_1=\frac{(YY^{\dagger})_{11}}{M_1}v_u^2\simeq 0.032\text{eV}\;.
\end{align}
Using the approximate efficiency function\cite{leptogenesis}
\begin{align}
\eta(m)=\left(\frac{8.25\times 10^{-3}\text{eV}}{m}+\left(\frac{m}{2\times 10^{-4}\text{eV}}\right)\right)^{-1}\;,
\end{align}
the predicted BAU in thermal leptogenesis scenario 
is given as
\begin{align}
\frac{n_B}{s}=-\frac{10}{31g_*}
\epsilon_1\eta(\tilde{m}_1)
=\begin{cases}
1.0\times 10^{-11}\left(\frac{m_M}{10^{13}\text{GeV}}\right)\;,
&\text{SM case}\;,\\
3.2\times 10^{-12}\left(\frac{m_M}{10^{13}\text{GeV}}\right)\;,
&\text{MSSM case}\;,\\
\end{cases}
\end{align}
with $g_* = 108.5$ in SM case and $g_*=232.5$ in MSSM case.
Note that $m_M$ is the only one parameter which are left free in 
our simplest model.
In order to reproduce $\eta_B/s = 8.7\times 10^{-11}$, 
we should set $m_M = 8.4\times 10^{13}$GeV for SM case and 
$m_M = 2.7\times 10^{14}$GeV for MSSM case which give
$(M_1, M_2, M_3)=
(3.1\times 10^{11}, 5.3\times 10^{12}, 8.9\times 10^{13})
\text{GeV}$ in SM case and 
$(M_1, M_2, M_3)=
(4.6\times 10^{11}, 7.9\times 10^{12}, 1.3\times 10^{14})
\text{GeV}$ in MSSM case.

\vskip 3mm 
\noindent
(c) The lepton flavor violation processes

It is interesting to consider $\mu\to e\gamma$ in the 
MSSM case. 
Even if the flavor universal boundary condition 
is taken at high energy scale such as GUT scale where 
all the sfermion mass matrices are proportional to $m_0^2\mathbf{1}$
with $\mathbf{1}$ being unit matrix and all the trilinear coupling 
matrix is 
proportional to the Yukawa coupling matrix with a dimensionfull 
proportionality coefficient $A_0$, the off-diagonal elements 
of neutrino Yukawa coupling matrix induce the flavor mixings in 
slepton sector and this affects the prediction on 
lepton flavor violating processes, such as $\mu\to e\gamma$.

In general, the SUSY contribution to $\mu\to e\gamma$ strongly 
depends on the right-handed neutrino mass scale 
in addition to the SUSY parameters.
The lower bound on the right-handed neutrino mass scale from 
successful leptogenesis 
\footnote{We don't consider the gravitino problem though
it is very serious problem\cite{gravitinoprob} 
in supersymmetric models. This topic is outside the scope of present work.} 
has an implication for the prediction of lepton flavor violation 
processes\cite{leptogenesisandlfv}.
In our simplest model, all the parameters are fixed.
Especially the successful leptogenesis gives $m_M \simeq 2.7\times 10^{14}$GeV.

The normalization factor for the right-handed neutrino mass matrix
$m_M = 2.7\times 10^{14}$GeV with $v_L = 3.0\times 10^{-11}$GeV
determines the normalization of neutrino Yukawa matrix as
$k_{\nu}=\sqrt{v_Lm_M/(v_u\lambda)^2}\simeq 2.1$.
With these normalization factors, one get
\begin{align}
Y^{\dagger}LY = 
\begin{pmatrix}
-0.0045&-0.0095-0.012i&-0.012+0.0040i\\
-0.0095+0.012i&-0.063&-0.013+0.038i\\
-0.012-0.0040i&-0.013-0.038i&-2.5
\end{pmatrix}
\end{align}
where 
$L=\mathrm{diag}(\ln\frac{M_1}{M_X}, \ln \frac{M_2}{M_X}, \ln \frac{M_3}{M_X})$ 
with $M_X=2.0\times 10^{16}$GeV and
we take the base where $M_R$ and $M_e$ are diagonalized. 
As easily seen from form of mass matrices, $m_D$, $M_R$, and $M_e$, 
large mixing angles in the neutrino sector come from seesaw 
enhancement\cite{SeesawEnhance}, 
{\it i.e.} the off-diagonal
elements of $m_D$ are not large even in the basis where $M_e$ and $M_R$ are
diagonalized.
Therefore the off-diagonal elements of $Y^{\dagger}LY$ is suppressed to be 
much smaller than one.

\begin{figure}
\begin{center}
\includegraphics[scale=2.0]{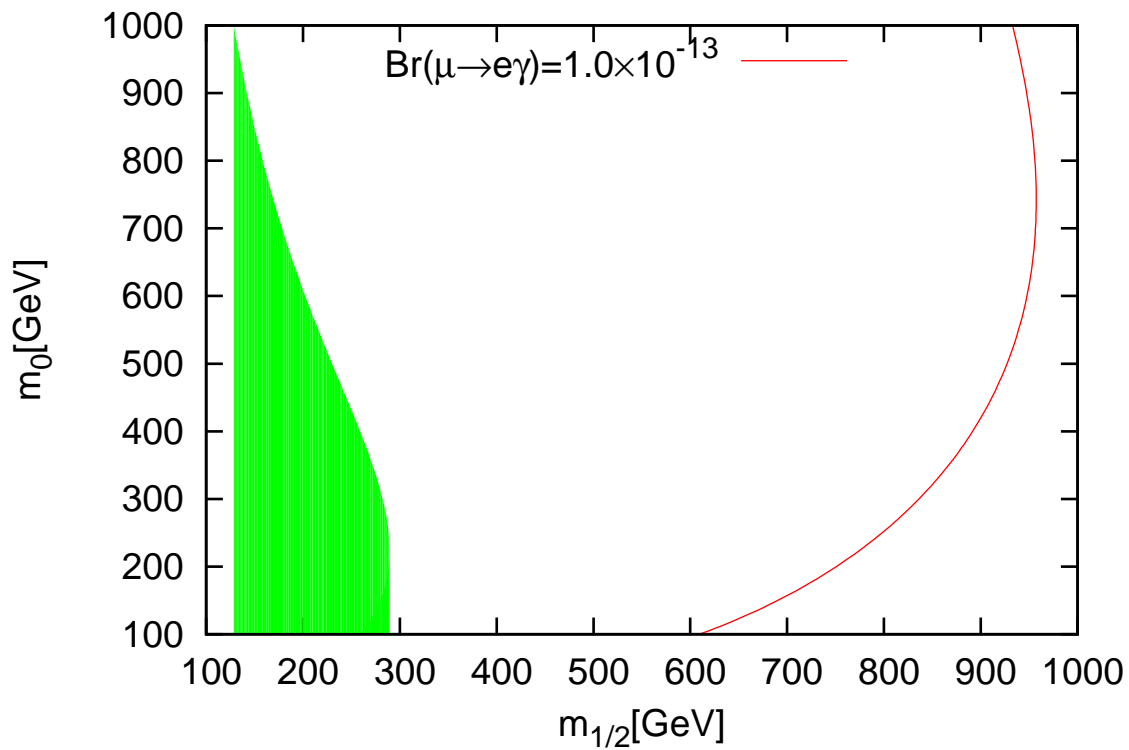}
\end{center}
\caption{The contour of predicted value of 
$\text{Br}(\mu\to e\gamma)=1.0\times 10^{-13}$ 
is shown on the $m_{1/2}$ and $m_0$ plane. 
The shaded region is already excluded, {\it i.e.}
$\text{Br}(\mu\to e\gamma)>1.2\times 10^{-11}$.
For the other SUSY parameters, 
$A_0=0$ and $\tan\beta=10$ are taken.}
\label{fig-mg-m0}
\end{figure}

The branching ratio of $\mu\to e\gamma$ 
within mass insertion approximation is calculated as
\begin{align}
\text{Br}(\mu\to e\gamma)\simeq\frac{\alpha^3}{G_F^2}
\frac{|6m_0^2+2A_0^2|^2}{(4\pi)^4m_S^8}|(Y^{\dagger}LY)_{12}|^2
\tan^2\beta\;.
\end{align}
If we take the grand unified gaugino mass, $m_{1/2}$ at $M_X$, 
$m_S^8$ is approximately $m_S^8=0.5m_0^2(m_0^2+0.6m_{1/2}^2)^2$\cite{PPTY}.
The $\mu\to e\gamma$ constraint on our model is displayed in 
Fig.~\ref{fig-mg-m0}. 
As shown in the figure, MEG experiment which is expected to 
reach $\text{Br}(\mu\to e\gamma)<10^{-13}$ can test very wide 
region of SUSY parameter space in our model.
A 3.4$\sigma$ deviation from the SM was reported in muon $g-2$\cite{expg-2}.
If this discrepancy comes from the SUSY contribution, 
rather light SUSY spectrum, {\it i.e.} $m_0, m_{1/2}<500$GeV,
is favored, so that our simple model promises a measurable size 
of $\text{Br}(\mu\to e\gamma)$ in the light of $g-2$.

\section{Summary and Discussions}
We examined whether neutrino mass matrix derived from the seesaw mechanism 
is compatible with the hierarchical mass matrices based on the Froggatt-Nielsen 
(FN) mechanism. We showed the followings:

\begin{enumerate}
	\item The milder hierarchy of neutrino masses 
is obtained for the FN charge of the left-handed 
lepton doublet $\ell_{Li}$ taken as $f_1=f_2+1=f_3+1$. 
Since the FN charges for $d_{Ri}$ are taken as (2,1,1),
this charge assignment of $\ell_{Li}$ is compatible with 
SU(5) GUT models with $f_1=2$.

\item We showed the problem which arises from the 2-3 symmetry 
is evaded by introducing two FN fields with opposite $Z_2$ parity. 
The relative phase of vacuum expectation values of two FN fields acts 
an important role for this.

\item We constructed a model where coupling constants are real. In this 
framework, the relative phase becomes the sole origin of complex phases 
in mass matrices. 
In particular, we examined a simple case where coupling matrices are 
proportional to the democratic matrix and obtained various predictions. 
The model predicts the normal hierarchy case and reproduces neutrino 
mixings, the ratio of neutrino squared masses well. The $\theta_{13}$, the 
Dirac CP phase and the two Majorana CP phases are predicted.  
The predicted effective mass of the double beta decay is small and 
the successful leptogenesis scenario is obtained. Also,  
the $\mu\to e+\gamma$ is discussed  in the SUSY model. 
\end{enumerate}

The intention of this paper is to show that the matching between 
the FN mechanism and the neutrino mass matrix derived from 
the seesaw mechanism is good, although 
the neutrino mass matrix must be quite different from that for the charged 
lepton mass matrix. 
Another interest is that this kind of 
model gives a possibility of the common origin of the CP violation for 
the quark system and the neutrino system, when we consider these mass 
matrices simultaneously. 

\vspace{1cm}
{\bf Acknowledgements.}  
It is a pleasure to thank Wilfried Buchm\"uller 
for useful discussions and comments.

\end{document}